\documentclass[
reprint,
superscriptaddress,
twocolumn,
amsmath,amssymb,
aps,
prl,
]{revtex4-2}

\usepackage{dcolumn}% Align table columns on decimal point
\usepackage{bm}% bold math
\usepackage{mathtools} % for equation alignment
\usepackage{color}  % for coloring edit command
\usepackage{comment}
\usepackage{braket}
\usepackage{tikz}
\usetikzlibrary{quantikz2}
\usepackage{float}
\usepackage{multirow}
\usepackage{siunitx}
\usetikzlibrary{snakes}
\usepackage{amsmath}
\usepackage{mhchem}
\usepackage{color,soul}
\usepackage[OT1, T1]{fontenc}
\DeclareTextSymbolDefault{\dh}{OT1}
\usepackage{booktabs}

\usepackage{graphicx}
\usepackage[caption=false, justification=centerlast]{subfig}
\usepackage[export]{adjustbox}

\usepackage{bm}
\usepackage{bbold}

\definecolor{mscolor}{rgb}{0,0.5,0.5}

\definecolor{tgcolor}{rgb}{0.5,0,0.5}

{}
\definecolor{phcolor}{rgb}{0.5,0,0.5}
\newcommand \be{\begin{equation}}
\newcommand \ee{\end{equation}}
\newcommand \bea{\begin{eqnarray}}
\newcommand \eea{\end{eqnarray}}
\newcommand \bse{\begin{subequations}}
\newcommand \ese{\end{subequations}}

\newcommand{\rsub}[1]{\textcolor{black}{#1}}

\newcommand{\UWM}{Department of Physics, University of Wisconsin-Madison, 1150 University Avenue, Madison, WI, 53706 USA}
\newcommand{\NBI}{Niels Bohr Institute, University of Copenhagen, Blegdamsvej 17, 2100 Copenhagen, Denmark}

\begin{document}

% indexing codes-- do these go in the manuscript somewhere?
% styleguide says i choose four.
% 32.80.Pj Optical cooling of atoms; trapping
% 33.80.P Optical cooling and trapping of atoms and molecules
% 32. Atomic spectra and interactions with photons
% 42. Optics

% \preprint{APS/123-QED}

\title{Efficient preparation of entangled states in cavity QED with Grover's algorithm}

\author{Omar Nagib}%
 \email{onagib@wisc.edu}
\affiliation{\UWM}
\author{M. Saffman}%
\affiliation{\UWM}
\author{K. M{\o}lmer}
\affiliation{\NBI}

\date{\today}% It is always \today, today,

\begin{abstract}
We propose to employ the amplification mechanism of Grover's search algorithm to efficiently prepare entangled states of an ensemble of qubits. The conditional change of sign employed in the algorithm can be implemented by the phase shift of photons scattered on an optical cavity hosting an atomic ensemble. We show that collective Dicke states, GHZ states, and Schr\"odinger cat superpositions of $N$ atoms may be prepared deterministically by few ($\sim N^{1/4}$) photon scattering events without individual addressing of the atoms. 

\end{abstract}

\maketitle

%\tableofcontents

\noindent
{\it Introduction. --}
There exist a host of methods to prepare  Dicke states  by use of   quantum circuits \cite{Bartschi2019Deterministic,Divide_Dicke_2022,Cruz_Dicke_2019,Kaye_Dicke_2001,Bartschi2019Deterministic,state_prep_shallow_circuits_2024,Cirac_Dicke_2024}, interactions \cite{Spin_squeeze_2024}, dissipation \cite{W_Dissipate_2017}, and measurements \cite{Dicke_phase_estimate_2021}. Such entangled states are useful resources for quantum sensing, communication and computing. While universal gate sets for quantum computing enable the preparation of any state from a trivial product state, qubit systems offering native, all-to-all interactions permit faster preparation of wide classes of permutation symmetric entangled states \cite{Sackett2000,Monz2009,Sorensen_Molmer_Ion_2000,Toffoli_superconducting_2011,Reed_2012,Rasmussen_2020,Sorensen_Klaus_1999,Bloch_2012,Molmer2011,Weimer_2010,Lanyon_2011,Bartschi_2022}. In this Letter, we propose a method to prepare entangled states of $N$ qubits that all interact with a single cavity mode in such a way that a photon reflected from the cavity experiences a phase shift of 0 or $\pi$, depending on the collective state of the qubits. Measuring such a conditional phase shift heralds the preparation of the desired target state and enables so-called carving schemes, proposed in
\cite{Sorensen2003probabilistic, Chen2015carving, Sorensen2003measurement, ramette2024counterfactual, Davis2018Painting} and implemented experimentally in \cite{Welte2017cavity, McConnell2015Entanglement, McConnell2013Generating, Christensen2014Quantum,Dordevic2021Entanglement, Florian2014Entangled}.
The carving schemes, however, only succeed with a probability given by the squared overlap of the initial state and the desired target state, which may be small in multi-qubit systems.

\begin{figure}
\includegraphics[width=.95\columnwidth]{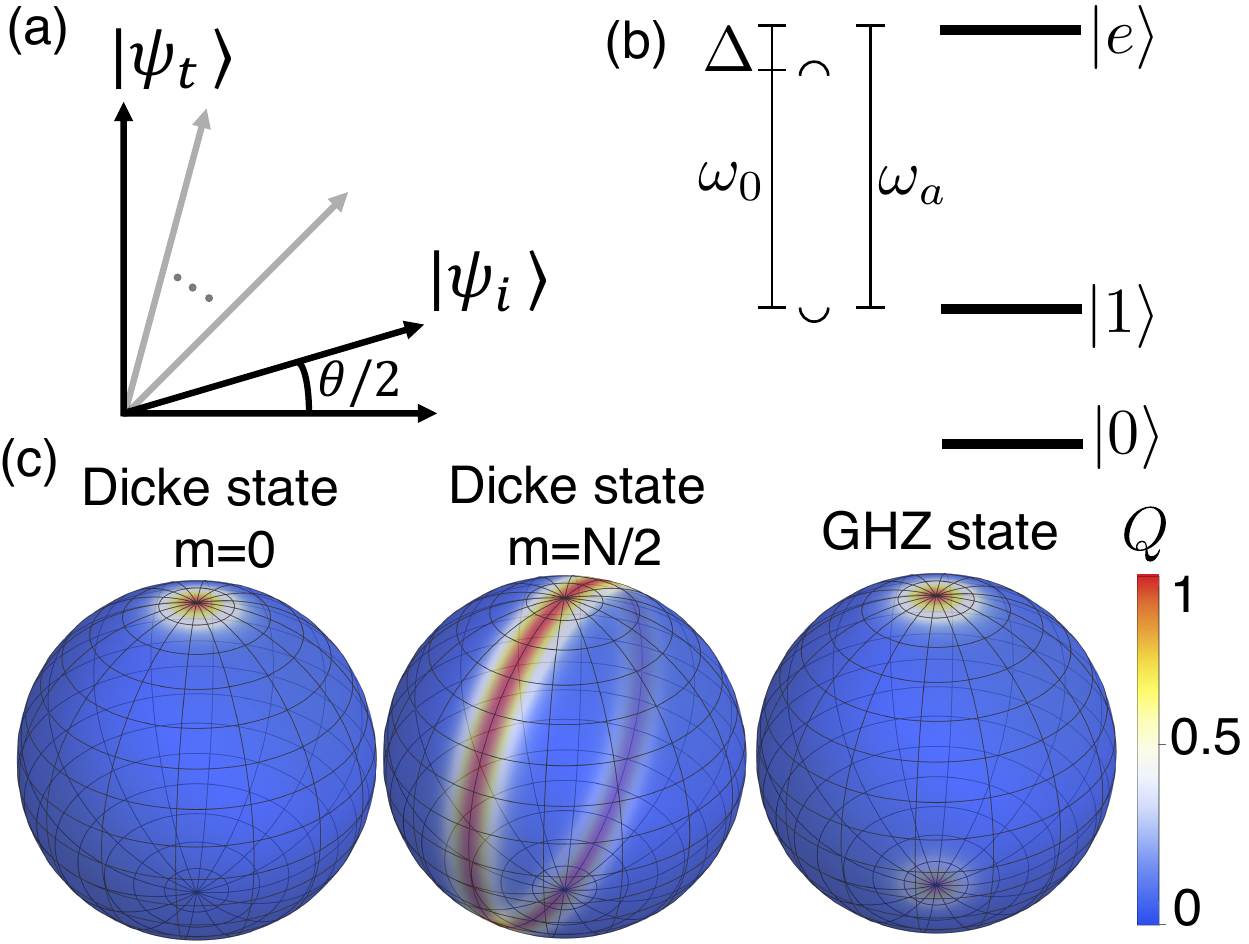}
    \caption{(a) \rsub{Each application of the Grover iteration rotates the initial state $\ket{\psi_i}$ towards the target $\ket{\psi_t}$ by an angle $\theta$}. (b) Energy level scheme for the atomic system interacting with the cavity field. (c) The Husimi Q-function for the ground atomic product state ($m=0$ Dicke state), the   $m=N/2$ Dicke state with respect to the spin-$y$ component and the GHZ state with $N=100$ qubits. The $Q$-function represents the overlap with the coherent spin product states (CSS). In each of the images, $Q$ is normalized by its maximum value. $Q(\phi,\beta)=\frac{3}{4 \pi}|\braket{\psi|\phi,\beta}^{\otimes N}|^2$.  }\label{fig1_PRL}
\end{figure}

\noindent
{\it Grover's algorithm. --}
Grover's search algorithm \cite{Grover1997} can be used to find an element that fulfills a certain criterion in an unstructured database starting from an initial product state that is a uniform superposition of all computational basis states.
Here we show that Grover's algorithm can also be used to deterministically and efficiently prepare an entangled target state $\ket{\psi_t}$ from an initial state $\ket{\psi_i}$ of qubits in a cavity, as long as both  of them can be identified by causing a change of sign in reflection of a single photon on the cavity.
The initial and final state must have a finite overlap $\sin(\theta/2) = \langle \psi_t\ket{\psi_i}$,
and the evolution progresses in the two-dimensional space spanned by $\ket{\psi_i}$ and $\ket{\psi_{t}}$. Defining $\ket{\psi_{t,\perp}}$ as orthogonal to $\ket{\psi_t}$ in this space, we can write:
\begin{equation}
 \ket{\psi_i}=   \sin(\theta/2) \ket{\psi_t} +  \cos(\theta/2) \ket{\psi_{t,\perp}}.
\end{equation}
Grover's algorithm applies the iteration of two successive unitary operations $G=\chi_{i}\chi_{t}$, where \rsub{$\chi_{t,i}= \mathbb{1}-2\ket{\psi_{t,i} }\bra{ \psi_{t,i} }$} puts a minus sign on the state component along $\ket{\psi_{t,i} }$ and leaves the other orthogonal state component unchanged.
In the Grover search algorithm, the initial state is simply a product state of all qubits in an equal superposition of states $\ket{0}$ and $\ket{1}$.  Following the selected phase shift of the target state, $\chi_i$ executes an inversion about the mean of all binary register state amplitudes, which increases the target state amplitude. Each step of the Grover algorithm can be visualized as an effective geometric rotation of the state vector in the direction from the initial $\ket{\psi_i}$ towards the target state  $\ket{\psi_t}$ by the angle $\theta$ [see Fig. \ref{fig1_PRL}(a)], and after $k$ applications of $G$ we obtain (up to a global phase),
\begin{equation}\label{eq:Gk}
G^k \ket{\psi_i}=  \sin \bigg[(2k+1)\frac{\theta}{2}\bigg] \ket{\psi_t} +  \cos\bigg[(2k+1)\frac{\theta}{2}\bigg]  \ket{\psi_{t,\perp}}.
\end{equation}
Since any classical register state has an overlap of $1/\sqrt{n}$ with the uniform superposition state in a Hilbert space of dimension $n=2^N$, it takes $k \propto \sqrt{n}$ operations to rotate the state through the angle $\sim \pi/2$ into the target state - this is the celebrated $\sqrt{n}$ Grover speedup over the classical search of a database with $n$ elements.  

\begin{figure}
  \includegraphics[width=0.45\textwidth]  {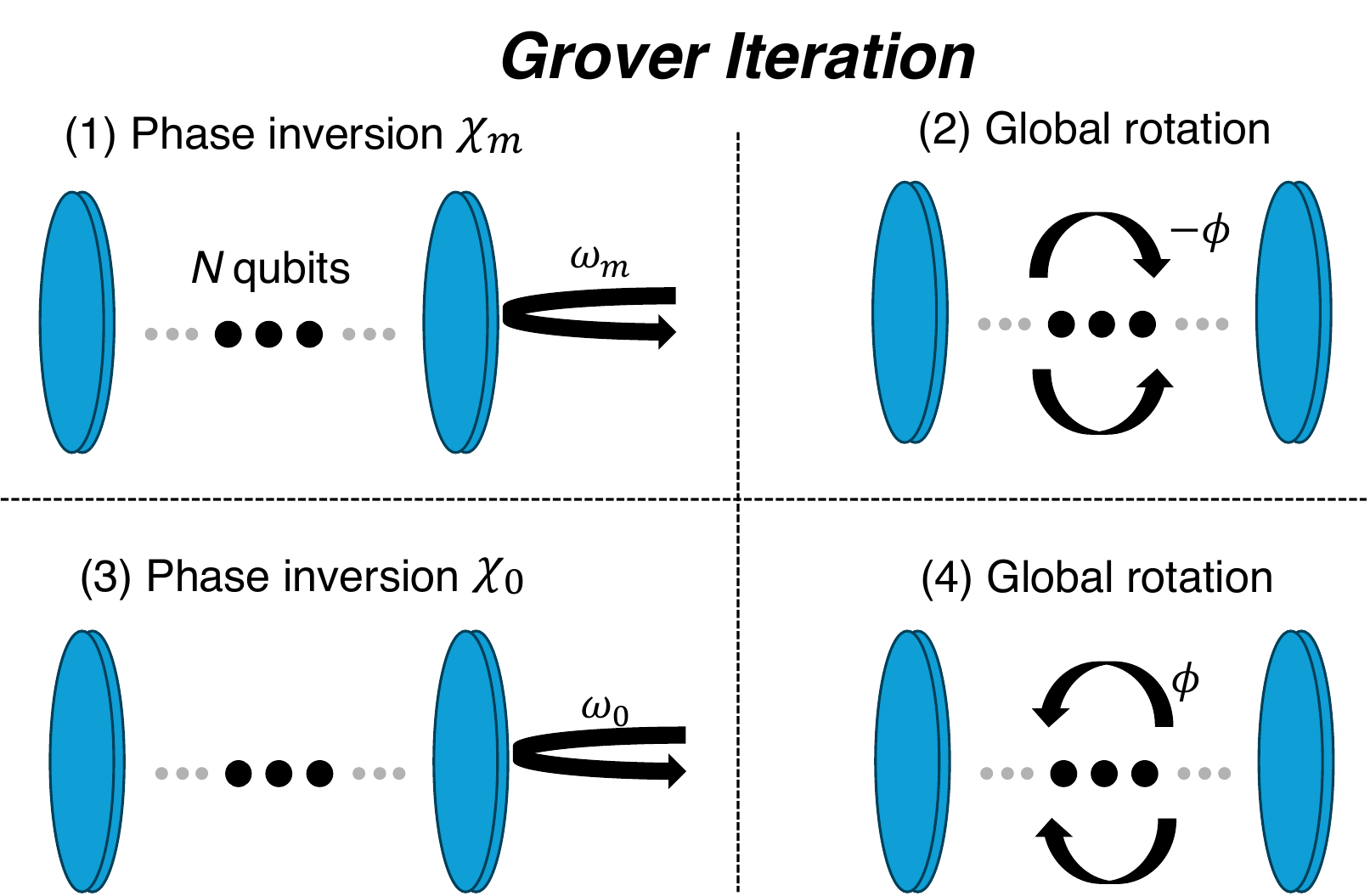}%
\caption{Physical implementation of the Grover iteration in cavity QED. Phase inversions in step (1) and (3) of the Dicke states $\ket{m=0}$ ($\chi_0$) and $\ket{m}(\chi_m)$ are realized by hitting the cavity with single photons with frequencies $\omega_0$ and $\omega_m=\omega_0+m\Omega$, respectively. Steps (2) and (4) are global spin rotations with opposite angles about the $y$-axis.}
\label{fig2_Grover_graphic2}
\end{figure}

The above geometric reasoning applies to the evolution between any two states, for which we can engineer the state-dependent phase shift operations, $\chi_i,\ \chi_t$. In this Letter we propose to use the Grover algorithm to prepare entangled states of ensembles of qubits by identifying initial states that are easy to prepare, easy to equip with the phase shifts given by $\chi_{i,t}$, and with sufficient overlap between the initial and target state to enable the state preparation in only a few steps. 

The state-dependent change of sign imposed by $\chi_{i},\chi_{t}$ in the Grover algorithm is conditioned on the state of every single qubit, and its implementation by one- and two-qubit gates in a quantum computer may be lengthy, and hence prone to decoherence and errors.  Efficient multi-qubit gates and schemes to implement the Grover algorithm by fewer physical operations have, however, been proposed for trapped ions \cite{XWang2000}, Rydberg atoms \cite{Molmer2011,Khazali2020, Keating2016Arbitrary,Grover_rydberg,Anikeeva2021}, and superconducting circuits \cite{Rasmussen2020}, employing effective all-to-all interactions in these systems. These schemes may hence also be applied to transfer an initial state into a final state and may be analyzed along the same lines as the photon-mediated proposal described below. 

\noindent
{\it State-dependent photon scattering. --} 
In this section, we show how the scattering phase shift of a single-photon wave packet on a one-sided cavity provides a deterministic protocol for the preparation of certain entangled states of $N$ atoms inside the cavity. Consider the energy-level scheme in Fig. \ref{fig1_PRL}(b): we assume the cavity has resonance frequency $\omega_0$ and the atoms have two ground states $\ket{0}$ and $\ket{1}$. The atomic transition $\ket{1} \leftrightarrow \ket{e}$ has resonance frequency $\omega_a$ that is coupled to the cavity mode with detuning $\Delta=\omega_0-\omega_a$, while $\ket{0}$ is far detuned and effectively uncoupled. The cavity detuning $\Delta$ from the atomic transition is much larger than the \rsub{atom-cavity coupling strength, $|\Delta| \gg g$}. Under these conditions, the atomic excited state can be eliminated and we recover a dispersive interaction Hamiltonian $H$ between the $N$ atoms and the light in the cavity \cite{Chen2015carving}:
\begin{equation}
H=\hbar \Omega \hat{m} \hat{n}_c, 
\end{equation}
where $\Omega=\dfrac{g^2}{\Delta}$, $\hat{n}_c$ is the photon number operator and $\hat{m}$ counts the number of atoms in state $\ket{1}$. 

We shall denote the symmetric eigenstates of $\hat{m}$ (the Dicke states) by $\ket{m}$. If the atoms occupy the state $\ket{m}$, the cavity resonance frequency shifts to the value $\omega_m=\omega_0+m\Omega$. Provided $\Omega$ is large enough, an incident single photon with frequency $\omega_m$ will be resonantly scattered and undergo a sign change if the atoms occupy the state $\ket{m}$, while it will be off-resonantly scattered with no change of sign if the atoms occupy any other Dicke state. Assuming that the photon leaves in the same wave packet state, its quantum state factors, and the scattering yields a deterministic phase gate on the atomic state $\ket{m'} \rightarrow e^{\imath\pi\delta_{m,m'}}\ket{m'}$, i.e. we implement $\chi_t$ with the target state $\ket{\psi_t}=\ket{m}$ \rsub{in a constant depth, independent of $N$}. We remark that this physical mechanism has been proposed previously to implement Grover's search algorithm to solve subset sum problems \cite{Anikeeva2021}.

\noindent
{\it Preparation of a Dicke state. --} 
It is easy to prepare an initial state with all atoms in $\ket{0}$, but this state is itself a Dicke state with $m=0$ and is orthogonal to the other Dicke states. We therefore   use as the  initial state a product state of all qubits in a rotated superposition  
\begin{eqnarray}\label{Psuc_Grover_Purify}
 \ket{\psi_i}&=&  \left[\cos (\phi/2) \ket{0}+\sin(\phi/2)\ket{1}\right]^{\otimes N}\nonumber \\
&=& \sum_{m=0}^{N} \sqrt{{N\choose m}} \cos^{N-m} (\phi/2) \sin^{m} (\phi/2) \ket{m}.
\end{eqnarray}
To put a change of sign on this state merely requires a coherent rotation of all atoms by angle $-\phi$, photon scattering with frequency $\omega_{m=0}$, and a final rotation of all atoms by the angle $\phi$, i.e., the operation $\chi_i$ is obtained as $R^{\otimes N}(\phi) \chi_{0} R^{\otimes N}(-\phi)$. Therefore, a single Grover step $G=\chi_i \chi_t$ to prepare the target Dicke state $\ket{m}$ would be given by
\begin{equation}\label{Grover_Dicke}
G=R^{\otimes N}(\phi)\chi_0R^{\otimes N}(-\phi)\chi_m
\end{equation}
i.e., it consists of two global rotations and photon reflections, independent of the qubit number and with no individual addressing required, as shown in Fig. \ref{fig2_Grover_graphic2}.

The number of Grover steps $k$ required to prepare a given Dicke state $\ket{m}$ from the optimally rotated coherent spin state (CSS) can be found by noting the value of the overlap, $\sin(\theta/2) = \sqrt{{N\choose m}} \cos^{N-m} (\phi/2) \sin^{m} (\phi/2)$. To prepare the Dicke state perfectly and rapidly in few integer steps, the rotation angle $\phi$ is chosen to satisfy the condition,
$\sin  \left[(2k+1)\theta/2\right]=1$, for the smallest integer $k$, cf., Eq. \eqref{eq:Gk}.

\begin{figure}[]
    \centering
    \includegraphics[width=0.45\textwidth]{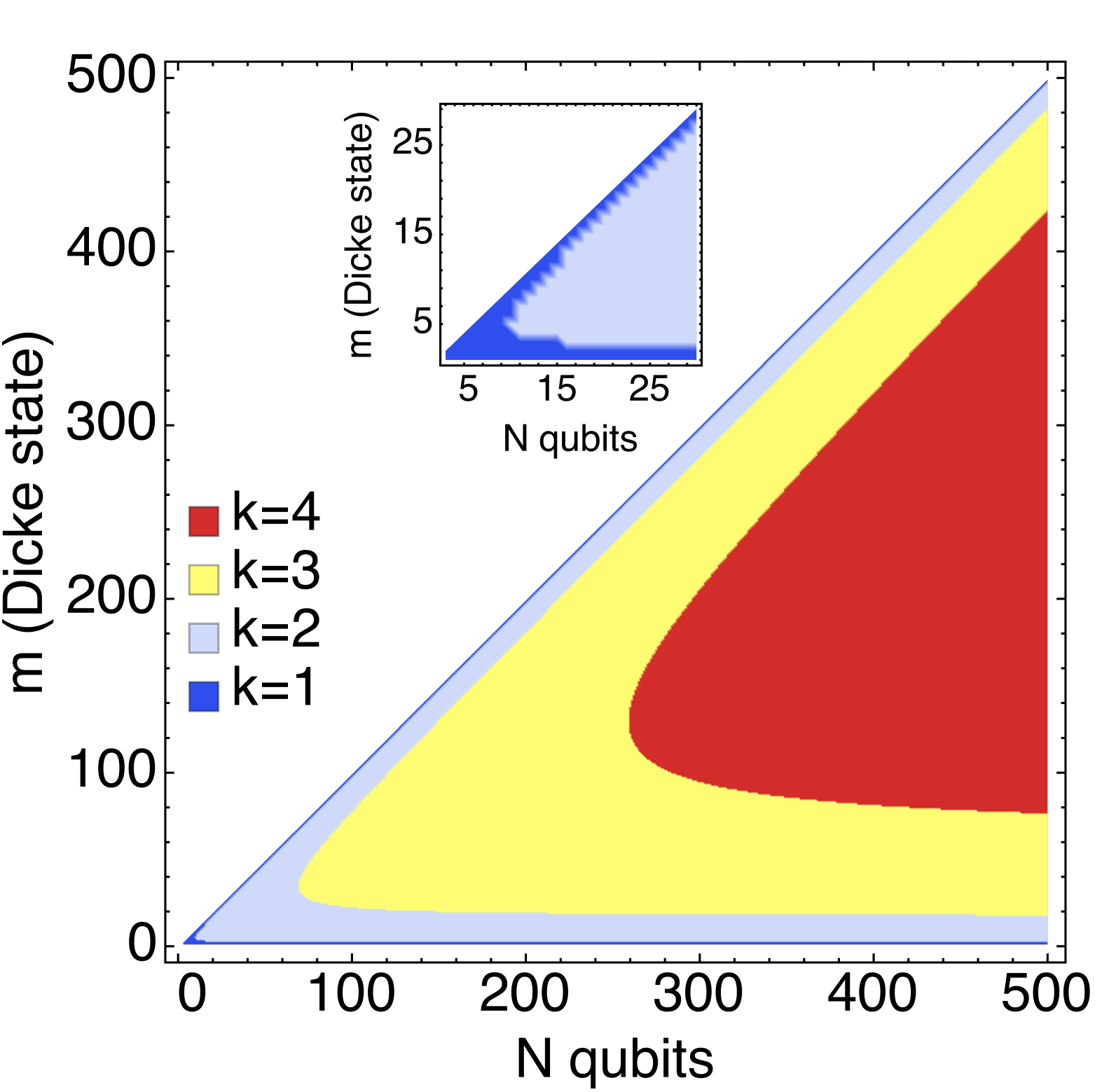}
    \caption{Contour plot showing the number of steps $k$ to prepare a Dicke state $\ket{m}$ with $N$ qubits.}
    \label{Fig:contour_plot}
\end{figure}

A simple estimate for $k$ is found by approximating the binomial factor in Eq. \eqref{Psuc_Grover_Purify} by a Gaussian with a variance ${\rm Var}(m)$ and a corresponding maximum value equal to $1/\sqrt{2\pi {\rm Var}(m)}$. For $m=N/2$, this yields $k \sim 0.88 N^{1/4} - 1/2$ where the subtracted constant ensures a near-perfect match with the numerical results. For large $N$ and finite values of $m$ the Poissonian limit of the binomial probability distribution yields a mean and variance equal to the desired value of $m$. For small $m$, the minimum number of Grover steps then depends on $m$ but is independent of $N$,
\begin{equation}\label{k_normal_modified}
 k \sim  1.24 m^{1/4}-\dfrac{1}{2}  . 
\end{equation}
In Fig. \ref{Fig:contour_plot}, we show a contour plot of the number of steps $k$ required to prepare the Dicke state $\ket{m}$ of $N$ qubits. Choosing the appropriate initially rotated CSS, we can prepare any Dicke state with qubit number $ 3\le N \le 500$ perfectly in four steps or less, i.e., by the scattering of 8 photons or less on the cavity. Dicke states with $m$ close to 0 or $N$ can be prepared in a single step, while states with $m\sim N/2$ are the most expensive. \rsub{In particular, the W state ($m=1$) can always be prepared in one step, independent of $N$}.

\noindent
{\it Preparation of a GHZ state. --} We proceed to show how Grover's algorithm can be used to prepare the GHZ state with a number of Grover steps $k = O(N^{1/4})$.
The state $\ket{\rm GHZ}=(\ket{0}^{\otimes N}+\ket{1}^{\otimes N})/\sqrt2$ is a superposition of Dicke states with $m=0$ and $m=N$. While we cannot put a phase change specifically on such a superposition by the scattering of a single photon, subsequent scattering of two photons, with frequencies $\omega_{m=0}$  and $\omega_{m=N}$, will put a minus sign on any superposition of these states. So, if the initial state has equal and non-vanishing overlap with the two extremal spin states, the Grover algorithm will amplify the GHZ state amplitude. 

We note that the initial state $\ket{m=0}$ has an overlap of $1/\sqrt{2}$ with the GHZ state, but it does not populate its components symmetrically and it will, in fact, stay invariant under the Grover steps. Rotating the state by $\phi=\pi/2$ yields identical amplitudes, but the state components with $m=0,N$ are in the wings of the binomial distribution and would require \rsub{exponentially} many Grover steps to become significantly populated.

Instead, we propose to first produce the Dicke state  $\ket{m=N/2}$ as described above and use it as a stepping stone for the process towards the GHZ state. Now, $\ket{m=N/2}$ has vanishing overlap with both Dicke state components of \rsub{the GHZ state}, but if we apply a global $- \phi$ rotation to all qubits, the rotated Dicke state overlaps the extremal states with equal amplitudes $\sim 1/N^{1/4}$, see Fig. \ref{fig1_PRL}(c). The Grover steps from the Dicke to the GHZ states thus take the form 
\begin{align}\label{Grover_GHZ}
&\ket{\psi_i}=R^{\otimes N}(-\phi)\ket{N/2}, \ \ \ket{\psi_t}=\ket{\rm GHZ}, \\ &  \nonumber G=R(-\phi)^{\otimes N}\chi_{N/2} R(\phi)^{\otimes N}\chi_0 \chi_N
\end{align}
where the sequence $\chi_{0} \chi_{N}$ puts the conditional phase on the target GHZ state, and $R(\pm \phi)^{\otimes N}$ rotates the initial Dicke state and rotates it back again for implementation of the conditional phase by $\chi_{N/2}$. As above, $\phi$ is chosen such that we end up with the GHZ state perfectly in the least number of integer steps. Note that the protocol employs three photon scatterings per iteration, but as the overlap between the initial and the two-component final state is larger by a factor of $\sqrt{2}$ than for the original preparation of the Dicke state, the resources are comparable to the ones needed to first prepare the Dicke state.

We may also wish to prepare Schr\"odinger Cat states, which are superpositions of spin coherent states of many qubits, with arbitrary different directions on the large spin Bloch sphere. In the companion paper \cite{Nagib2025b} to this letter, we show that a similar protocol to that of the GHZ state can be applied to prepare such Cat states if the components are sufficiently separated.

\begin{table}
  \centering
\rsub{
\begin{ruledtabular}
\begin{tabular}{lcc}
  Scheme (Ref.) & Depth / Resource & Infidelity \\[2pt] \colrule
  \multicolumn{3}{l}{\textbf{Algorithms}} \\[2pt]
  Dicke $m$: Ref. \cite{Cirac_Dicke_2024}                    & $O(m^{1/4}\ell_{m,\epsilon}^{2})$ & $\epsilon$ \\
  \phantom{Dicke $m$: }Ref. \cite{yu2024efficientpreparationdickestates} & $O(\ln m+\ln1/\epsilon)$         & $\epsilon$ \\
  \phantom{Dicke $m$: }Present                                & $O(m^{1/4})$*                     & 0          \\[4pt]
  GHZ $N$: Ref. \cite{QC_LOCC}                                & $O(1)$                            & 0          \\
  \phantom{GHZ $N$: }Ref. \cite{Cruz_Dicke_2019}              & $O(\ln N)$                        & 0          \\
  \phantom{GHZ $N$: }Present                                  & $O(N^{1/4})$*                     & 0          \\[6pt] \midrule
  \multicolumn{3}{l}{\textbf{Implementations}} \\[2pt]
  Cavity carving \cite{Chen2015carving}                       & $O(m^{1/2})$** photons              & $O(C^{-1})$ \\
  Grover-based \cite{Grover_geometric_2020}                 & $O(N^{5/4})$ phase gates          & $O(1-e^{-\pi^{2}\kappa/g})$ \\
  Grover-based \cite{Grover_2023}                          & $O(N)$ phase gates                & N/A         \\
  Ref. \cite{yu2024efficientpreparationdickestates}             & $O(\ln m)$ photons***             & N/A         \\
  Present (unherald)                                           & $O(m^{1/4})$ photons              & $O(C^{-1/2})$ \\[2pt]
   Present (herald)                                           & $O(m^{1/4})$ photons              & $O(C^{-2/3})$ \\[2pt]
\end{tabular}
\end{ruledtabular}
\caption{\label{tab:summary}
Comparison of algorithms and physical implementations for preparing Dicke and GHZ states. $\ell_{m,\epsilon}
= \log_{2}\{(1/\ln(4/3))\,[2m(\ln(2m)+9/2)
  + \ln(\mathrm{Poly}(m)/\epsilon^{2})]\}\,$. *Assumes a constant‑depth Grover step; **the number of repetitions needed for the carving scheme to succeed; ***each photon carries $O(N)$ frequency components.}}
\end{table}
\noindent
{\it Error analysis. --} 
\rsub{Given that single-qubit global rotations can be experimentally executed with very high fidelity ($\ge 99.9\%$) \cite{Lukin_global_rotation_2023,Muniz2025}, the main error contribution in the Grover unitary [Eqs. \eqref{Grover_Dicke} and \eqref{Grover_GHZ}] is due to the phase inversion $\chi_m$. The fidelity of $\chi_m$ is fundamentally limited by the ratio of the coherent atom-cavity interaction to dissipation as quantified by the cooperativity $C=g^2/\kappa \gamma$, where $\kappa$ ($\gamma$) is the cavity (atom) decay rate. Here, we give a simplified physical argument for the scaling of the infidelity with $C$, with a more detailed analysis of the photon scattering process deferred to  the companion article \cite{Nagib2025b}. We first note that the cavity is able to spectrally distinguish two neighboring Dicke states $\ket{m}$ and $\ket{m\pm 1}$ when the cavity resonance shift $\Omega=g^2/\Delta$ is much greater than the cavity linewidth $\kappa$, i.e., we require $d=\Omega/\kappa \gg1$. The infidelity due to a finite cavity ``resolution'' scales as $1/d^2=(\kappa/\Omega)^2$, since $\Omega=\omega_{m+1}-\omega_m$ is the detuning between the resonance frequencies of the two neighboring Dicke states, and the lineshape of a cavity is Lorentzian-like with a tail decaying quadratically in the detuning. A second source of error comes from spontaneous emission, which for $m$ atoms coupled dispersively to the cavity occurs at an effective rate $\gamma (g/\Delta)^2$. Taking the interaction time of the photon with the cavity as $t\sim 1/\kappa$, the probability of spontaneous emission is then $\sim m\gamma (g/\Delta)^2 t = m\gamma g^2/\Delta^2 \kappa$ or $m d^2/C$. Therefore, the infidelity equals the sum of the two errors $1-F \sim 1/d^2 +md^2/C$, which attains a minimum at $d\sim (C/m)^{1/4}$ leading to the  scaling}\rsub{
\begin{equation}
1-F(\chi_m) \sim \dfrac{1}{\sqrt{C}}\ 
\end{equation}}
\rsub{While this holds for $m \neq 0$, it can be shown that $1-F(\chi_0) \sim 1/C$ for $m=0$ \cite{Nagib2025b}. Moreover, the infidelity can be reduced further to $1-F(\chi_m)\sim C^{-2/3}$ by heralding on detection of the reflected photon \cite{Nagib2025b}.} 

\rsub{Another nonideality arises from the frequency dependence of the scattering amplitudes, where photon pulses of finite bandwidth $\sigma$ acquire slightly different temporal shapes and, hence, the factoring of the atomic state and the photon wave packet is not exact. After reflection, the atomic and photonic states are entangled as $\sum_{n} c_n \ket{n}\otimes \ket{\psi_n}_p$, and tracing over the photonic qubit leads to reduced coherence and infidelity. The reduced coherence is related to the overlap between the corresponding photonic states as ${\rm Re}(\braket{\psi_n|\psi_l})\sim{\rm Re}\{\int d\omega |\Phi(\omega)|^2  r_n(\omega) r^*_l(\omega)\}$, where $|\Phi(\omega)|^2$ is the photon wavepacket and $r_n(\omega)$ is the reflection amplitude of the photon when there are $n$ atoms coupled to the cavity. The wavepacket is taken to be a Gaussian $e^{ -(\omega-\Omega_c)^2/2 \sigma^2}/\sqrt{2 \pi} \sigma$ with a bandwidth $\sigma$ and a central frequency $\Omega_c=mg^2/\Delta$. For a sufficiently narrow wavepacket, $\sigma/\kappa \ll1$, centered around $\Omega_c$, we can Taylor expand the integrand as $  r_n(\omega) r^*_l(\omega) \sim r_n(\Omega_c)r_l^*(\Omega_c)+a(\omega-\Omega_c)+b(\omega-\Omega_c)^2$, where $a$ and $b$ are constants. Since the wavepacket is a Gaussian that is symmetric around $\Omega_c$, the integration of the first-order term vanishes while the second-order term yields $\sigma^2$. Therefore, the leading order infidelity contribution is $1-F \sim (\sigma/\kappa)^2$.}

\begin{figure}[!t]
    \centering
    \includegraphics[width=0.47\textwidth]{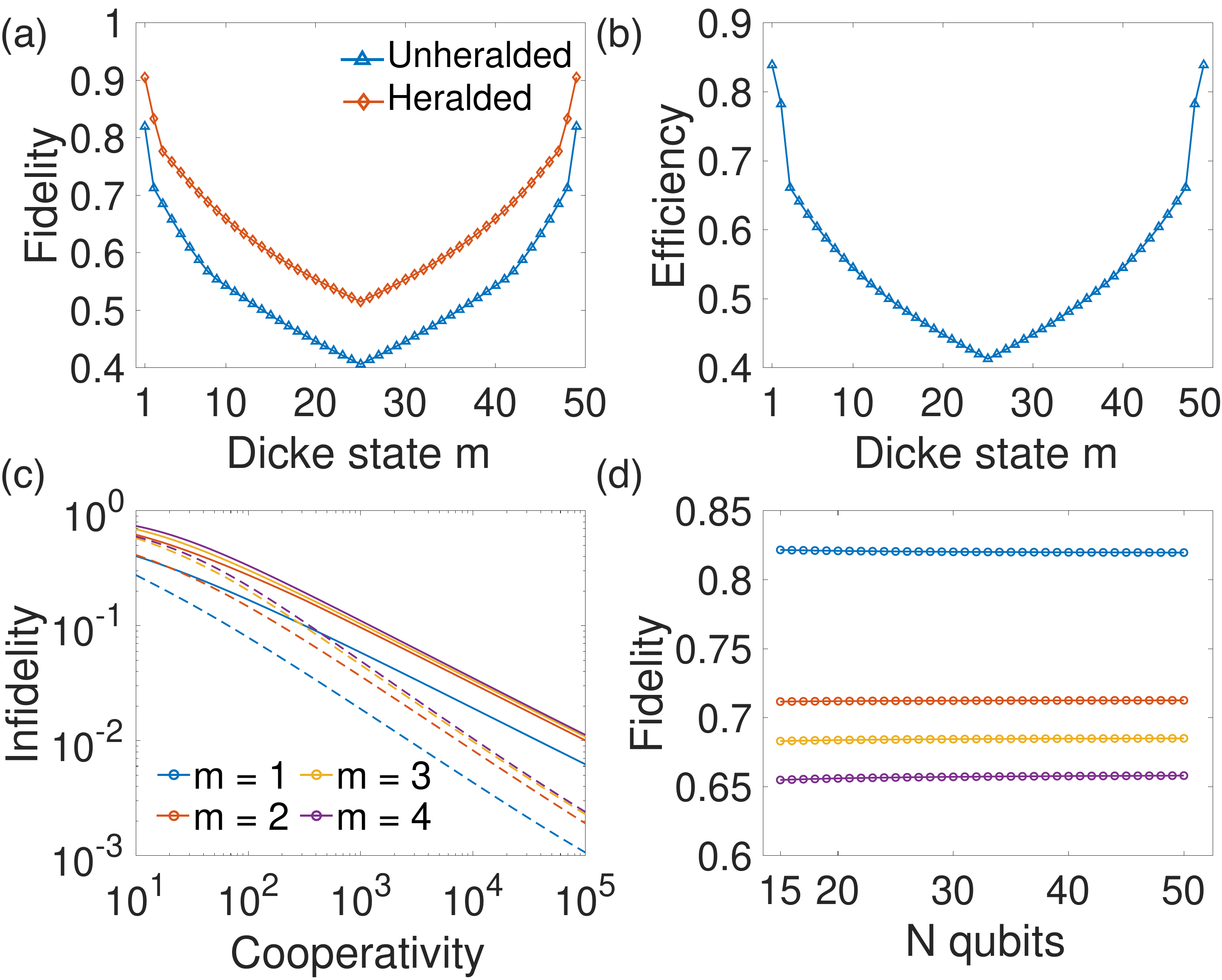}
    \caption{(a) Fidelity of Dicke states  $\ket{m}$ generated by Grover's algorithm without (blue) and with (red) heralding by detection of the reflected photon. The system parameters used are $g=10, \ \kappa=1, \gamma=1, \ C=100,$ and the photon wave packet is Gaussian with a frequency spread $\sigma=0.1 \kappa$. Internal and mirror losses are not included. Both $\Delta$ and $k$ have been optimized to maximize the fidelity. (b) The success probability of the protocol with heralding. (c) Scaling with the atom-cavity cooperativity of the infidelity of Dicke states for $N=15$ atoms, for the cases of not heralding (solid) and heralding (dashed). The systems parameters are as in panels (a) and (b) but the photons wave packet has $\sigma=0.01\kappa$. (d) For small $m$-values, the fidelity of the Dicke states is almost independent of the number of qubits. We show unheralded fidelities for the same system parameters as in panels (a) and (b). }
    \label{fig4_output_figure}
\end{figure}

\rsub{Fig. \ref{fig4_output_figure}(a) shows the results of numerical simulation of the fidelity in preparing various Dicke states for realistic parameters for current optical cavities with $C=10^2$ \cite{High_C}. The highest fidelity is limited to $70-80\%$ for Dicke states with the smallest $m$ (after which it decreases steeply), which may be boosted to $80-90\%$ by heralding, at the expense of finite success probability $\sim 80\%$ [Fig. \ref{fig4_output_figure}(b)]. Plotting the infidelity versus $C$, Fig. \ref{fig4_output_figure}(c) shows that achieving a fidelity of $99 \%$ would require $C> 10^4$ and $C> 10^5$ for the cases of heralding and not heralding, respectively. One advantage of the present scheme is that for $m \ll N$, the fidelity to prepare the Dicke state $\ket{m}$ is roughly independent of the qubit number [Fig. \ref{fig4_output_figure}(d)]. In ultra-low-loss nanofiber Fabry-Perot cavities, cooperativity of up to $C\sim 2 \times10^3$ has been reported \cite{Ruddell_2020}, which would give a maximum fidelity of $90-97\%$ for the Dicke states $m \in [25,1]$ with heralding. }

\noindent
{\it Outlook. --}
In summary, we have presented a method to prepare several different entangled states of $N$ qubits in a cavity employing the amplitude amplification mechanism of the Grover search algorithm, and single photon scattering events for the individual Grover steps. We provide protocols for the efficient preparation of Dicke states, GHZ states and Cat states, and we anticipate that a much wider range of states and operations can be engineered by similar mechanisms \cite{garratt2024quantumalgorithmpreparequasistationary}. We provide, elsewhere \cite{Nagib2025b}, a detailed analysis of the fidelity of the protocols.

Table \ref{tab:summary} summarizes the \rsub{resources} needed for preparing $N$-qubit Dicke states \rsub{and GHZ states and compares with previous work}. By using Grover's algorithm, the present work achieves a quadratic improvement over previous probabilistic cavity carving schemes that require $O(m^{1/2})$ attempts \cite{Chen2015carving}. \rsub{While Ref. \cite{yu2024efficientpreparationdickestates} achieves a better scaling of $O(\log m)$ trial steps, using repeated collective measurements and feedforward, we note that the actual number of steps achieved by the present algorithm is smaller as a function of $N$ and $m$ (cf. Fig. 2(d) in \cite{yu2024efficientpreparationdickestates} and Fig. \ref{Fig:contour_plot})}. The present results may also be compared with a recent proposal, where a quantum circuit assisted with ancillas, midcircuit measurements, and feedforward prepares Dicke states with circuit depth $O(m^{1/4})$ (up to a polylogarithmic correction), using Grover's amplitude amplification \cite{Cirac_Dicke_2024}. Our results have comparable scaling, without invoking ancillas, individual addressing, and midcircuit measurements.

Previous proposals employing atom-cavity interactions and Grover's algorithm for Dicke state preparation require resources that scale (super) linearly with the qubit number \cite{Grover_geometric_2020,Grover_2023}. In contrast, the resources for the physical implementation of the phase gate do not scale with the qubit number in the present work, and thus only $O(m^{1/4})$ applications of the phase gates and one rotation angle are needed for state preparation.

\rsub{The present proposal can be implemented in physical systems  where an ancilla qubit couples to the system qubits operator $\hat{m}$. This occurs naturally in many other platforms such as Circuit QED \cite{Circuit_QED_2009}, trapped ions \cite{ions_QED_2007}, and Rydberg atoms \cite{Molmer2011,Khazali2020, Keating2016Arbitrary,Anikeeva2021}. In the Rydberg platform, an atomic ancilla in the Rydberg state replaces the photon, and the Rydberg blockade mechanism serves the same goal as the photon blockade in the cavity QED implementation \cite{Anikeeva2021}.}

A physical demonstration of our protocols is realistic for a few tens of qubits given current experimental capabilities with high cooperativity cavity-mediated atom-photon interactions.  In order to scale to hundreds of qubits with good fidelity we require cavity cooperativity above $10^3-10^4$, which is challenging at optical frequencies. This limitation may inspire implementation of our protocols in \rsub{other platforms with very strong interaction, such as Rydberg atoms \cite{Anikeeva2021} or superconducting qubits, where $C=10^3-10^5$ is achievable \cite{Niemczyk2011,rollano2022highcooperativitycouplingnuclear} and cooperativity as large as $10^9$ has been reported \cite{Chakram_2021}}. 
       
This material is based upon work supported by the U.S. Department of Energy Office of Science National Quantum Information Science Research Centers as part of the Q-NEXT center, as well as support from NSF Grant No. 2016136 for the QLCI center Hybrid Quantum Architectures and Networks, and NSF Grant No. 2228725.   KM acknowledges funding from the Carlsberg Foundation through the “Semper Ardens” Research Project QCooL and from the Danish National Research Foundation (Center of Excellence “Hy-Q,” grant number DNRF139).\\

\bibliography{carve,qc_refs,rydberg}

\end{document}